\documentclass[journal]{IEEEtran}
\usepackage{graphicx,subcaption}
\usepackage{cite}
\usepackage{url}     
\usepackage{soul}
\usepackage{verbatim}
\usepackage{times}		
\usepackage{multirow}	
\usepackage [justification=centering]{caption}
\usepackage{amsmath}
\usepackage{amsfonts}
\usepackage{lipsum}
\usepackage{mathtools}
\usepackage{cuted}
\usepackage{hyphenat}	
\usepackage{lineno} 
\usepackage[usenames]{color} 
\usepackage{textcomp} 
\usepackage[dvipsnames]{xcolor}
\definecolor{myblue}{rgb}{0.180, 0.453, 0.683}
\usepackage{comment}
\usepackage{pk_math_am}
\usepackage{pk_math_nz}

\newcommand{\degree}{\ensuremath{{^\circ}}\xspace}

\newcommand{\fF}{\ensuremath{\mathit{fF}}\xspace}

\newcommand{\reviseJSSC}[1]{{\bf \color{myblue} #1}}

\renewcommand{\reviseJSSC}[1]{}

\begin{document}
\raggedbottom

\title{Amplitude-to-Phase Conversion in Injection-Locked CMOS Ring Oscillators}
%
\author{	Zhaowen~Wang\vspace{-2em}
}%

\maketitle
%
\begin{abstract}
  Injection-locked ring oscillators~(ILROs) are extensively employed for multi-phase clock generation in wireline and optical links. However, existing injection-locking theories primarily rely on linearized phase-domain or nonlinear time-domain models, which fail to account for amplitude-to-phase conversion effects inherent in ILROs. This paper introduces an enhanced analytical model based on an extension of Adler’s equation, explicitly incorporating amplitude-to-phase conversion. Simulation results demonstrate strong alignment with the proposed analytical predictions, validating the model’s accuracy in capturing the locking range and phasor relationships.  
\end{abstract}
%
%
\section{Introduction}
\label{sec:intro}

\IEEEPARstart{M}{ulti}-phase clock generation finds many critical applications in high-speed communication systems, including multi-phase sampling, multi-phase receiver, digital transmitter, phase interpolation, phase array, frequency synthesis, and so on~\cite{Wang2022_thesis,Chen2018,Poon2021,Wang2022,Xu2024,Li2024,Wang2024,Wang2022JSSC}. Injection-locked ring oscillators (ILROs) are widely used for multi-phase clock generation in high-speed wireline and wireless communications due to their symmetric structure, low jitter, and compact size~\cite{Kinget2002,Wang2021}. CMOS ILROs are particularly advantageous over their current-mode-logic (CML) counterparts due to their wider tuning range and lower jitter~\cite{Abidi2006,Wang2021JSSC}. However, a theoretical understanding of the injection lock mechanism in ILRO remains elusive.

Conventional injection-locking analyses include frequency-domain, time-domain, and phasor-domain approaches. Frequency-domain analysis, exemplified by Adler's equation~\cite{Adler1946,Razavi2004,Yang2019}, was originally developed for harmonic oscillators, such as LC oscillators, where a resonator tank filters out harmonics and maintains a dominant fundamental tone, allowing for linear superposition. This method is unsuitable for ring oscillators, which lack such a tank and exhibit highly nonlinear waveforms.

Time-domain analysis models each ring oscillator stage as a linear RC filter with a hard-limiting transconductance~\cite{Gangasani2006,Gangasani2008}. While this captures some transconductance nonlinearity, it neglects amplitude dependency. Furthermore, it better suits CML-based ROs, as CMOS ROs lack explicit RC filters.

Phase-domain response~(PDR) analysis uses the impulse sensitivity function to model phase noise and locking range~\cite{Hajimiri1998, Hong2019}. While PDR captures the overall injection effect, it does not distinguish individual RO stages or analyze phase imbalance. Moreover, as a simulation-based approach, it offers limited design insights.

Several analytical models have been proposed for injection-locked ROs, but they are either too simplified to capture the amplitude dependency~\cite{Raj2015} or too complex to provide clear design insights~\cite{Zhang2022}.

This paper presents an extended Adler's equation, a phasor-based analysis that incorporates amplitude dependency in CMOS ROs through amplitude-to-phase coefficients. Because CMOS ROs operate at high frequencies, their waveforms are still predominantly characterized by the fundamental frequency, making a phasor-based approach a reasonable approximation. The analytical results from the proposed model show good agreement with simulations, validating its effectiveness. This model provides new insights into the locking range, phase balance, and overall operation of ILROs.

The paper is organized as follows: Section~\ref{sec:RO Model} describes the CMOS RO model. Section~\ref{sec:adler's} presents the extended model that incorporates amplitude to phase and compares the analytical and simulation results. Section~\ref{sec:conclusions}  concludes the paper.

\section{CMOS Ring Oscillator Modeling}
\label{sec:RO Model}

\begin{figure}[t!]
\centering 
\includegraphics[width=0.45\textwidth]{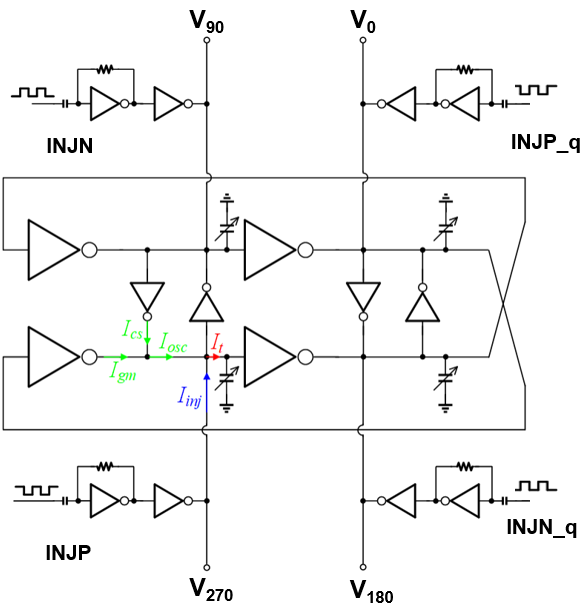}
\vspace{-6pt}
\caption{\small{The schematic diagram of an injection-locked ring oscillator with multi-phase injection.}}
\label{fig:ro}
\vspace{-12pt}
\end{figure}

Even-stage CMOS ring oscillators (ROs) are commonly used for generating pseudo-differential clocks, so they will be the primary focus of this paper. Each stage employs a pair of cross-coupled inverters to ensure differential operation and suppress common-mode DC gain, preventing RO deadlock~(Fig.~\ref{fig:ro}). Our RO model begins with a detailed modeling of a single stage.

A single RO stage comprises a differential input pair and a cross-coupled pair~(Fig.~\ref{fig:ro}). In a two-stage differential RO, the input to the first stage is 0~\degree and 180~\degree, producing outputs at 270~\degree and 90~\degree, respectively. The cross-coupled pair is driven by these differential signals (e.g., 90~\degree and 270~\degree in the first stage). The main inverter generates a current phasor $I_{gm}$ from $V_{0}$, while the cross-coupled pair contributes a current phasor $I_{cs}$ driven by $V_{90}$. The summation of these currents, i.e. $I_{rosc}$,  results in a phase shift from $I_{gm}$, which is defined as $\theta_{VI}$. In the two-stage differential RO, each stage needs to provide 90~\degree phase shift, which is impossible through the conversion of current to voltage by first-order RC filtering. Thanks to the extra phase shift of $\theta_{VI}$, the current-to-voltage conversion is less than 90~\degree, which fulfills the oscillation requirement.

\begin{figure}[t!]
\centering 
\includegraphics[width=0.45\textwidth]{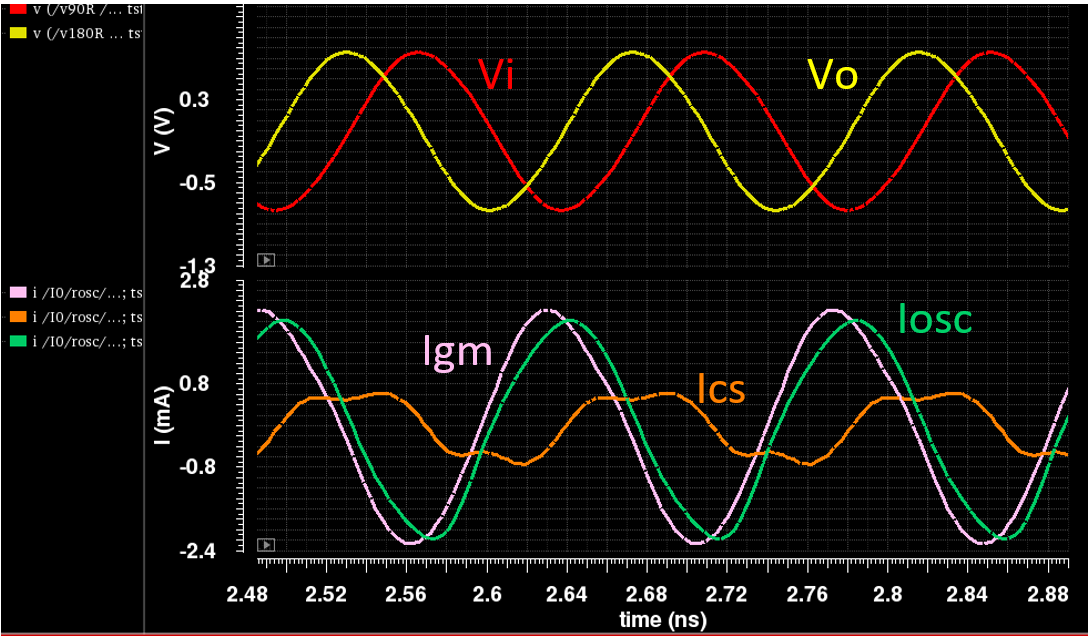}
\vspace{-6pt}
\caption{\small{The waveform of a 7-GHz free-running ring oscillator.}}
\label{fig:wave}
\vspace{-12pt}
\end{figure}

\begin{figure}[t!]
\centering 
\includegraphics[width=0.45\textwidth]{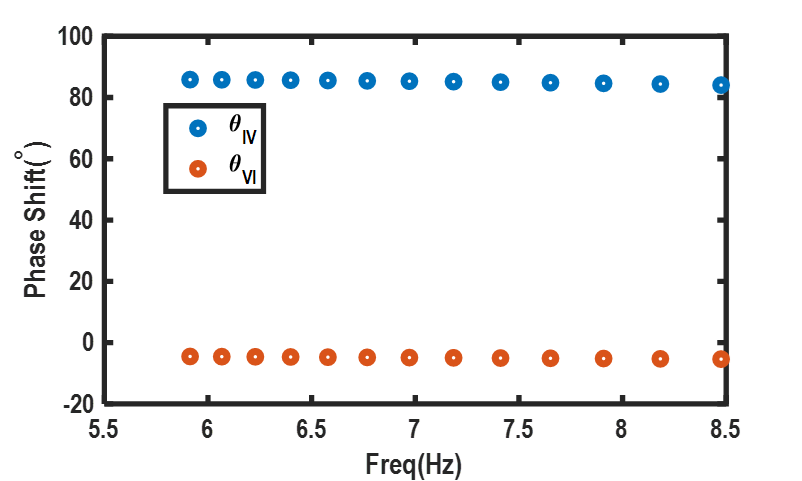}
\vspace{-6pt}
\caption{\small{$\theta_{IV}$ and $\theta_{VI}$ versus the frequency of a free-running ring oscillator.}}
\label{fig:theta_fr}
\vspace{-12pt}
\end{figure}

Fig.~\ref{fig:wave} shows the voltage and current waveforms of a free-running two-stage differential RO at 7GHz. Due to the high-frequency operation, the voltage waveform approximates a sinusoid. The main inverter, operating similarly to a class-AB amplifier, converts the input voltage into current. While it behaves as a hard current limiter at low speeds, it exhibits unity voltage gain in steady-state high-speed, high-swing operation. Its output current $I_{gm}$ is predominantly the fundamental frequency. Moreover, the total current $I_{osc}$ is shifted by the cross-coupled pair current $I_{cs}$, and it is still dominated by the fundamental tone. The current phase shift $\theta_{VI}$ is almost close to -5~\degree, and the current-to-voltage conversion is around 85~\degree, so the total single-stage phase shift is 90~\degree. $\theta_{VI}$  and $\theta_{IV}$ are almost constant when the free-running frequency varies from 6~GHz to 8.5~GHz as shown by Fig.~\ref{fig:theta_fr}. Because the oscillation waveform is mostly dominated by the fundamental tone, a phasor-based analysis will be valid. In an eight-stage ring oscillator, the voltage waveform is closer to square waveform, and contains more high-order harmonics. Nevertheless, as the frequency is higher, its feedforward branch boosts the free-running frequency and als make its waveform more close to sinusoidal~\cite{Sun2019}. Therefore, in high-speed applications, phasor analysis still applies to 8-stage ROs.

\begin{figure}[t!]
\centering 
\includegraphics[width=0.45\textwidth]{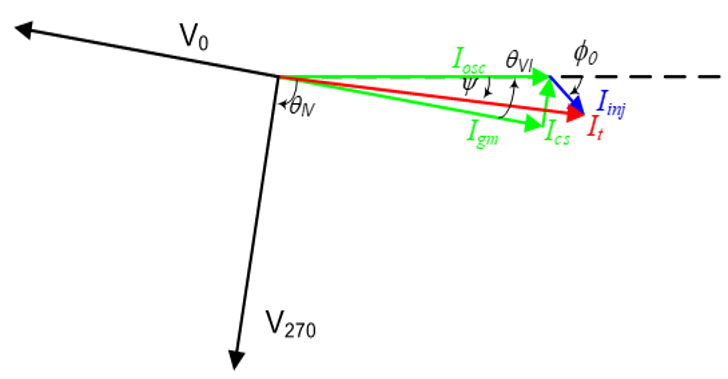}
\vspace{-6pt}
\caption{\small{The phasor diagram of the voltages and currents in an injection-locked ring oscillator.}}
\label{fig:phasor}
\vspace{-12pt}
\end{figure}

Fig.~\ref{fig:phasor} shows the phasor diagram in an RO. $I_{gm}$ is the current driven by $V_0$, $\theta_{VI}$ is the phase shift from the cross-coupled pair current $I_{cs}$.
With injection locking, the injection current $I_{inj}$ flows into the oscillation node, introduces an extra current summation and phase shift, and locks the RO to the injection frequency. In a multi-phase injection-locked RO, each oscillation node has its phase-shifted injection signals. $\psi$ is the phase shift due to injection current $I_{inj}$, $\theta_{IV}$ is the phase shift from the current-to-voltage conversion, and the total current $I_t$ generate the next-stage voltage $V_{270}$. $\phi_0$ denotes the angle between $I_{inj}$ and $I_{osc}$. In a multi-phase injection-locked case, $V_0$ and $V_{90}$ have the same amplitude and have a quadrature relationship, and so do $I_{t,0}$ and $I_{t,90}$.

\begin{figure}[t!]
\centering 
\includegraphics[width=0.4\textwidth]{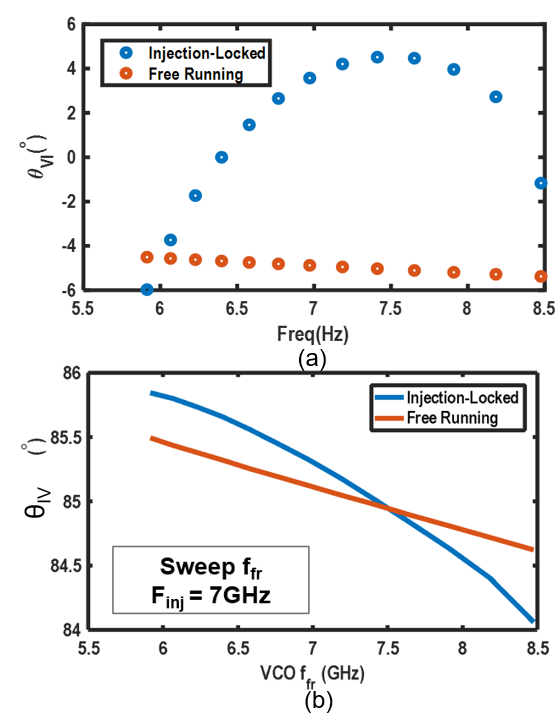}
\vspace{-6pt}
\caption{\small{$\theta_{IV}$ and $\theta_{VI}$ versus the free-running frequency of a ring oscillator with multi-phase injection enabled and disabled.}}
\label{fig:theta_ilo}
\vspace{-12pt}
\end{figure}

By sweeping the capacitor from 80~\fF to 124~\fF in the ring oscillator, its free-running frequency varies from 6~GHz to 8.5~GHz. Both $\theta_{IV}$ and $\theta_{VI}$ are independent from the free-running frequency, indicating a consistent behavior in the ring oscillator (Fig.~\ref{fig:theta_ilo}). However, as the multi-phase injection enabled, only $\theta_{VI}$ varies against the free-running frequency, while $\theta_{IV}$ barely changes. Therefore, the angle of current-to-phase conversion $\theta_{IV}$ is almost a constant regardless of the injection conditions.

\begin{figure}[t!]
\centering 
\includegraphics[width=0.4\textwidth]{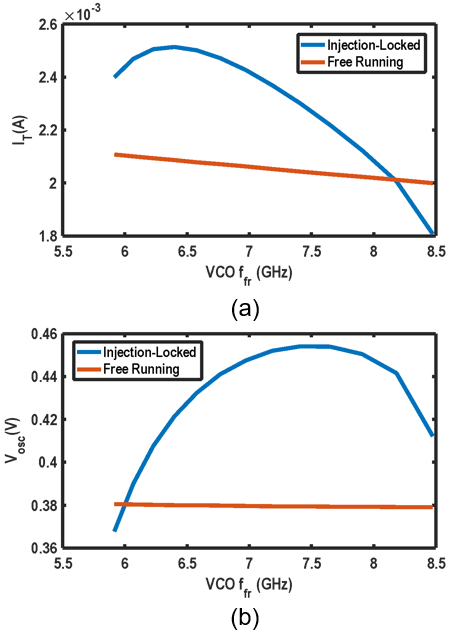}
\vspace{-6pt}
\caption{\small{the phasor amplitude $I_{t}$ and $V_{osc}$ in the RO with multi-phase injection enabled and disabled.}}
\label{fig:amp}
\vspace{-12pt}
\end{figure}

Fig.~\ref{fig:amp} further shows the phasor amplitude $I_{t}$ and $V_{osc}$ of the free-running and multi-phase injection locked RO. The free-running RO has a quite stable oscillation current and voltage amplitude while the injection-locked RO has a varying amplitude for either current or voltage. Moreover, the current reaches its maximum near the lower end of the locking range, while the voltage reaches its maximum near the higher end of the locking range.

Despite their nonlinear relationship with frequency, the voltage and current amplitudes exhibit a largely linear relationship, as shown in Fig.~\ref{fig:amp}. Since the injection is typically designed to be between 0.05 and 0.2 of the main stage, it can be treated as a small signal. Consequently, the amplitude disturbance caused by the injection remains minimal. For instance, with an injection strength of 0.2, the amplitude varies only between 0.37~V and 0.45~V, representing approximately $20\%$. Thus, small-signal analysis can be effectively applied to study the sensitivity of the main stage and the cross-coupled pair~\cite{Auvergne2000,Kabbani2003}. These components can be modeled as a combined transconductance stage with a phase shift, where the amplitude is represented by $I_{rosc}$ and the phase shift by $\theta_{VI}$. Using linear extrapolation, we derive the following equations, where  $A_{VI}$ and $G_{m}$ are the current phase and amplitude sensitivity to its input voltage amplitude:

\begin{equation}
\label{eqa:amp1}
\begin{split}
& \theta_{VI}=A_{VI}|\overrightarrow{V_{in}}|+ \theta_{VI,0} 
\end{split}
\end{equation}
\begin{equation}
\label{eqa:amp2}
\begin{split}
& |\overrightarrow{I_{osc}}|=G_{m}|\overrightarrow{V_{in}}|+ I_{osc,0} 
\end{split}
\end{equation}

\begin{figure}[t!]
\centering 
\includegraphics[width=0.4\textwidth]{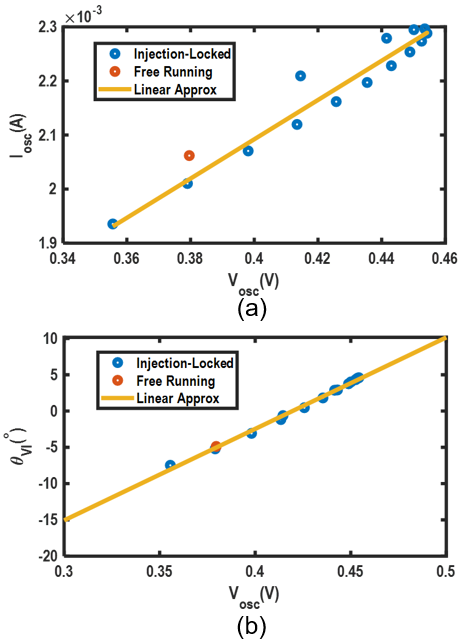}
\vspace{-6pt}
\caption{\small{The amplitude and phase of $I_{osc}$ in the RO with multi-phase injection enabled and disabled.}}
\label{fig:extra}
\vspace{-12pt}
\end{figure}

For the RO, the current-to-votlage delay of each stage can be expressed by $T_d=\theta_IV/(2\pi f)$. Alternatively, this delay can also be estimated by considering the time required to charge the total loading capacitance. In the steady state, the current and amplitude across the capacitor must satisfy the relationship $T_d = k*(CV_{osc})/I_t$, where $k$ is a fixed coefficient accounting for the conversion from sinusoidal charging to linear current charging. From this, we can derive $k=I_t\times\theta_{IV}/(2\pi fV_{osc}C)$. Since $k$ is a constant as shown by~Fig.~\ref{fig:k}, the following equation holds true for oscillators both with and without injection:

\begin{equation}
\begin{split}
& \frac{|I_{osc,fr}|\theta_{IV,fr}}{C|V_{osc,fr}|f_{fr}} = \frac{|I_{t}|\theta_{IV,inj}}{C|V_{osc,inj}|f_{inj}} 
\end{split}
\end{equation}

\begin{figure}[t!]
\centering 
\includegraphics[width=0.4\textwidth]{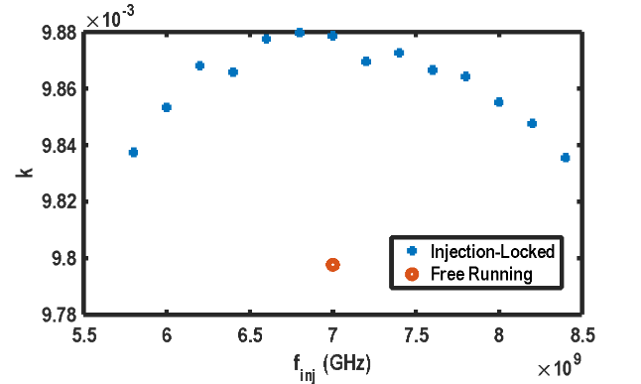}
\vspace{-6pt}
\caption{\small{k versus different injection frequencies.}}
\label{fig:k}
\vspace{-12pt}
\end{figure}

We now have a model for the single stage in the ring oscillator (RO) that accounts for its amplitude dependency through the amplitude-to-amplitude coefficient $G_m$ and amplitude-to-phase coefficient $A_{VI}$. This model can be seamlessly incorporated into Adler's equation.

\section{Extended Adler’s Equation with Amplitude-to-Phase Conversion}
\label{sec:adler's}
\subsection{Adler's Equation}
\label{sec:adler's eqa}

Adler's equation describes injection locking by analyzing the summation of current phasors~\cite{Adler1946}. The oscillator's current phasor is $I_{osc}$, and the injection current phasor is $I_{inj}$. The total current is shifted by an angle of $\psi$ from $I_{osc}$ due to the influence of $I_{inj}$. In the conventional analysis, When $I_t$ is orthogonal to $I_{inj}$, $\psi$ reaches its maximum at $\arctan(I_{inj}/I_{osc})$. This relationship is instrumental in analyzing the locking range, noise sensitivity, and other characteristics of an injection-locked oscillator. For an N-stage multi-phase injection-locked RO, Adler's equation can be expressed as:
\begin{equation}
\begin{split}
& \overrightarrow{I_{osc}}+ \overrightarrow{I_{inj}} = \overrightarrow{I_{t}}
\end{split}
\end{equation}
\begin{equation}
\begin{split}
& \theta_{VI} + \psi - \theta_{IV} = \pi/N
\end{split}
\end{equation}
\begin{equation}
\begin{split}
& \frac{\theta_{VI,fr}}{f_{fr}} =\frac{\theta_{VI,inj}}{f_{inj}} 
\end{split}
\end{equation}

\begin{figure}[h]
\centering 
\includegraphics[width=0.4\textwidth]{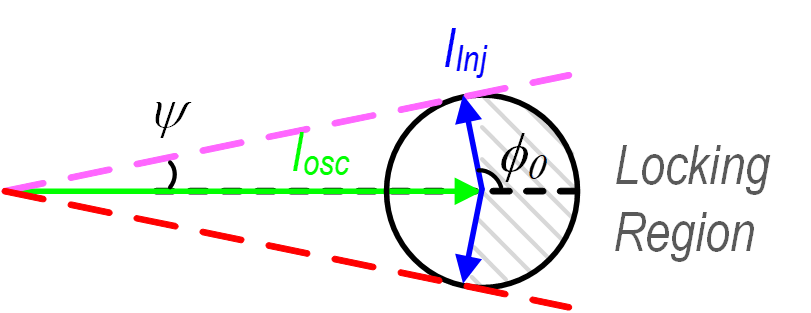}
\vspace{-6pt}
\caption{\small{Phasor diagram for Adler's equation.}}
\label{fig:k}
\vspace{-12pt}
\end{figure}

\subsection{Extended Adler's Equation}
\label{sec:ext adler's eqa}
To extend Adler's equation by incorporating amplitude-to-phase conversion, Equations~\ref{eqa:amp1} and \ref{eqa:amp2} are introduced. Injection locking alters the total current amplitude, necessitating adjustments to the model. As a result, the complete set of equations can be expressed as:
\begin{equation}
\begin{split}
& \overrightarrow{I_{osc}}+ \overrightarrow{I_{inj}} = \overrightarrow{I_{t}}
\end{split}
\end{equation}
\begin{equation}
\begin{split}
& \theta_{VI} + \psi - \theta_{IV} = \pi/N
\end{split}
\end{equation}
\begin{equation}
\begin{split}
& \frac{|\overrightarrow{I_{osc,fr}}|\theta_{VI,fr}}{\overrightarrow{V_{osc,fr}}f_{fr}} =\frac{|\overrightarrow{I_{t}}|\theta_{VI,inj}}{\overrightarrow{V_{osc}}f_{inj}}
\end{split}
\end{equation}
\begin{equation}
\begin{split}
& \theta_{VI}=A_{VI}|\overrightarrow{V_{in}}|+ \theta_{VI,0} 
\end{split}
\end{equation}
\begin{equation}
\begin{split}
& |\overrightarrow{I_{osc}}|=G_{m}|\overrightarrow{V_{in}}|+ I_{osc,0} 
\end{split}
\end{equation}
\begin{equation}
\begin{split}
& \phi_{0} = -\angle \overrightarrow{I_{inj}}
\end{split}
\end{equation}
\begin{equation}
\begin{split}
& \psi = -\angle \overrightarrow{I_{t}}
\end{split}
\end{equation}

\begin{figure}[t!]
\centering 
\includegraphics[width=0.4\textwidth]{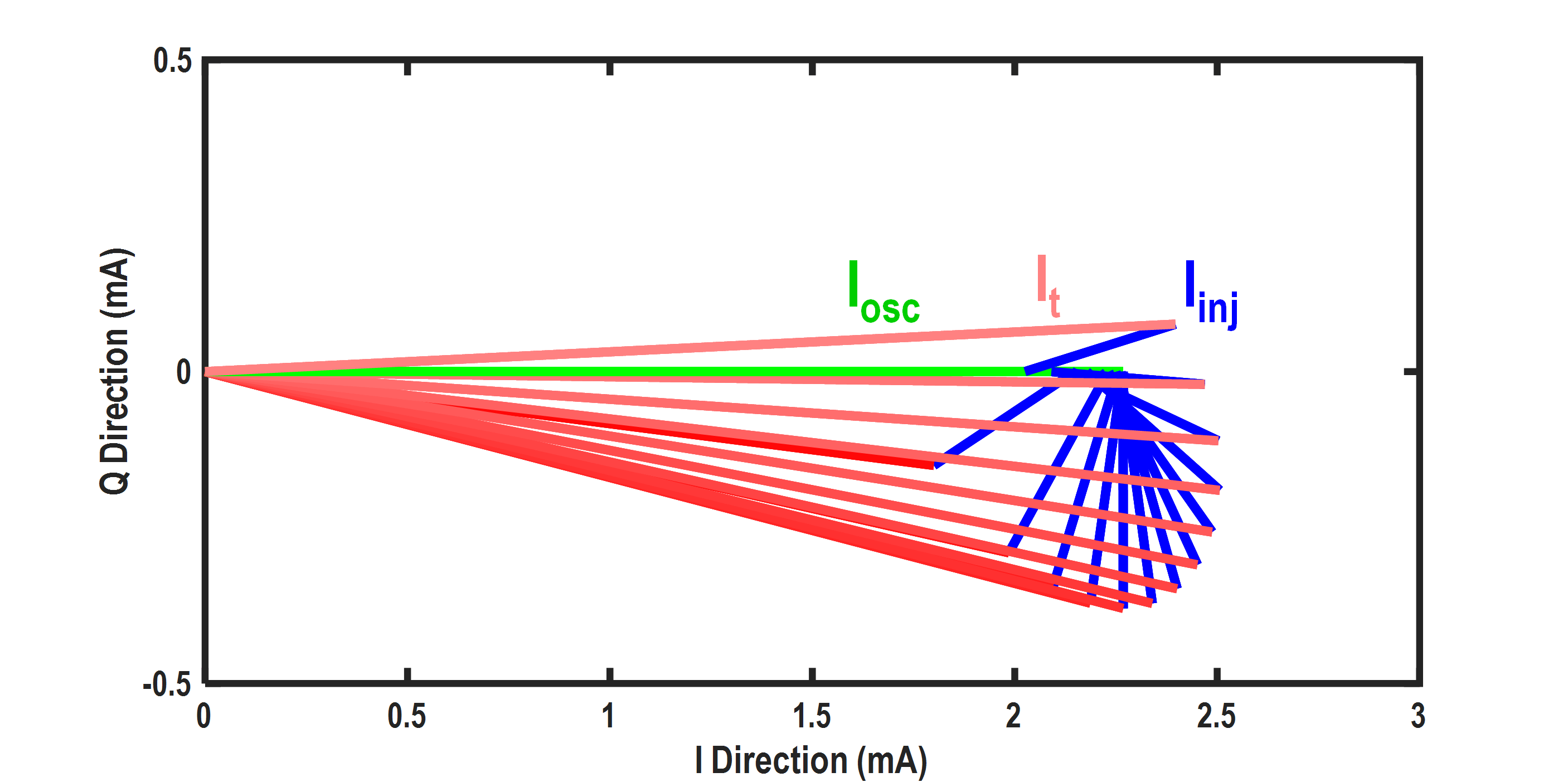}
\vspace{-6pt}
\caption{\small{Phasor diagram for the injection-locked RO with amplitude-to-phase conversion.}}
\label{fig:Fig_IL_ROSC}
\vspace{-12pt}
\end{figure}

Unlike the symmetric locking range predicted by the conventional Adler's equation, the extended equation accounts for amplitude-to-phase conversion, resulting in an asymmetric locking range. Additionally, the phasor diagram demonstrates a varying amplitude, as illustrated in Fig.~\ref{fig:Fig_IL_ROSC}.

\subsection{Analytical and Simulation Results}
\label{sec:results}
We examine the equations using a two-stage multi-phase injection-locked ring oscillator.

\begin{figure}[t!]
\centering 
\includegraphics[width=0.48\textwidth]{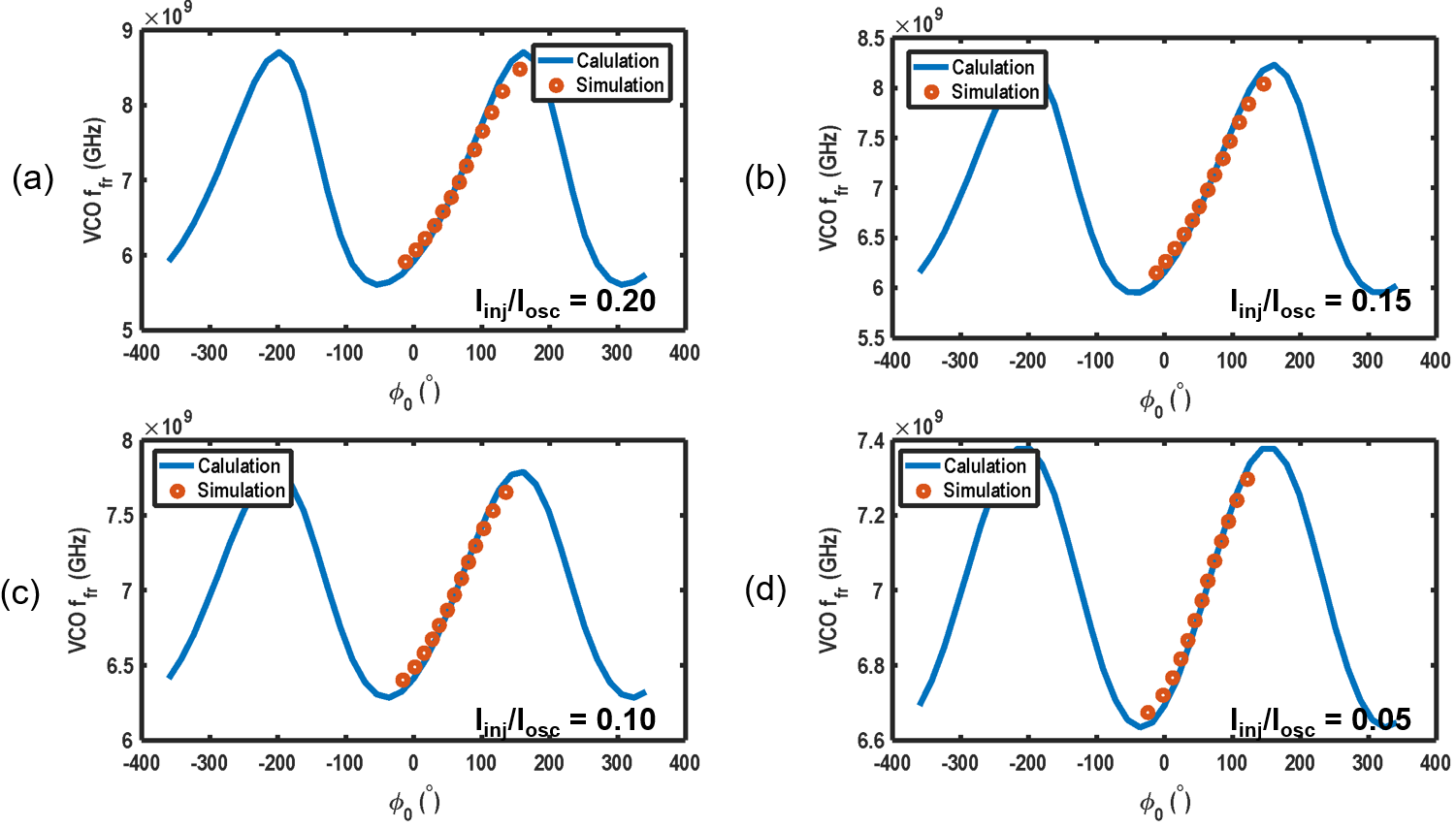}
\vspace{-6pt}
\caption{\small{Calculated and simulated $\phi_0$ versus free-running frequency under injection strength: (a) 0.2; (b) 0.15; (c) 0.10; (d) 0.05.}}
\label{fig:sw_ffr_phi0}
\vspace{-12pt}
\end{figure}

\begin{figure}[t!]
\centering 
\includegraphics[width=0.48\textwidth]{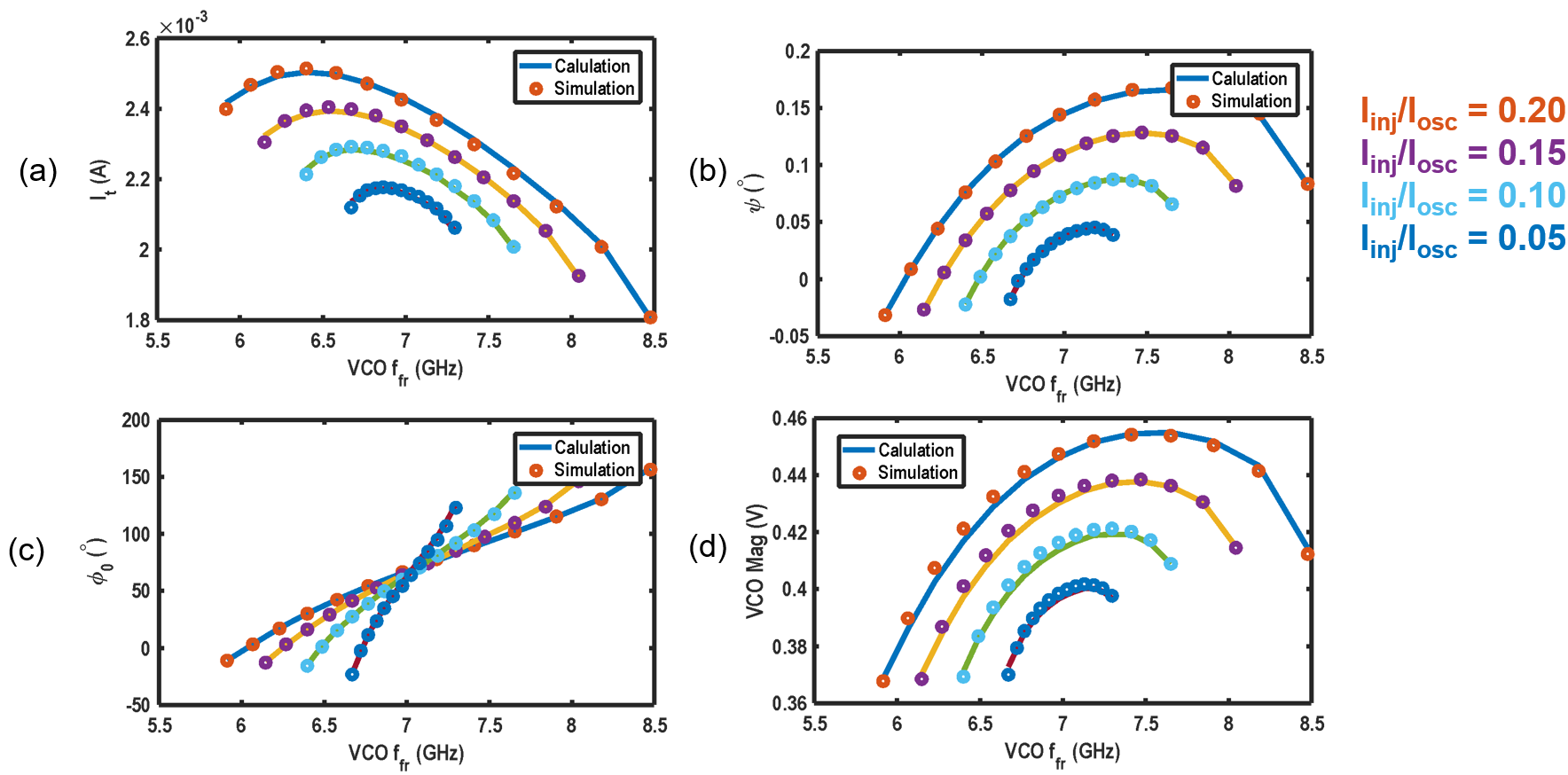}
\vspace{-6pt}
\caption{\small{Calculated and simulated $I_t$,$\psi$,$\phi_0$,$V_{osc}$ versus free-running frequency under injection strength:(a) 0.2; (b) 0.15; (c) 0.10; (d) 0.05.}}
\label{fig:sw_ffr_phi0_var}
\vspace{-12pt}
\end{figure}

First, the injection frequency is fixed at 7 GHz, and the free-running frequency is swept under varying injection strengths of 0.05, 0.10, 0.15, and 0.20. As shown in Fig.~\ref{fig:sw_ffr_phi0}, the equations provide analytical solutions for $\phi_0$, which correspond to specific locking ranges for each injection strength. The results indicate a wider locking range at the high-frequency end and a narrower range at the low-frequency end. Fig.~\ref{fig:sw_ffr_phi0_var} compares the current amplitude, $\phi_0$ and $\psi$, and oscillation amplitude across the free-running frequency for different injection strengths. The extended equations accurately predict these values, validating the amplitude-to-phase conversion effect in injection-locked oscillators.


\begin{figure}[t!]
\centering 
\includegraphics[width=0.48\textwidth]{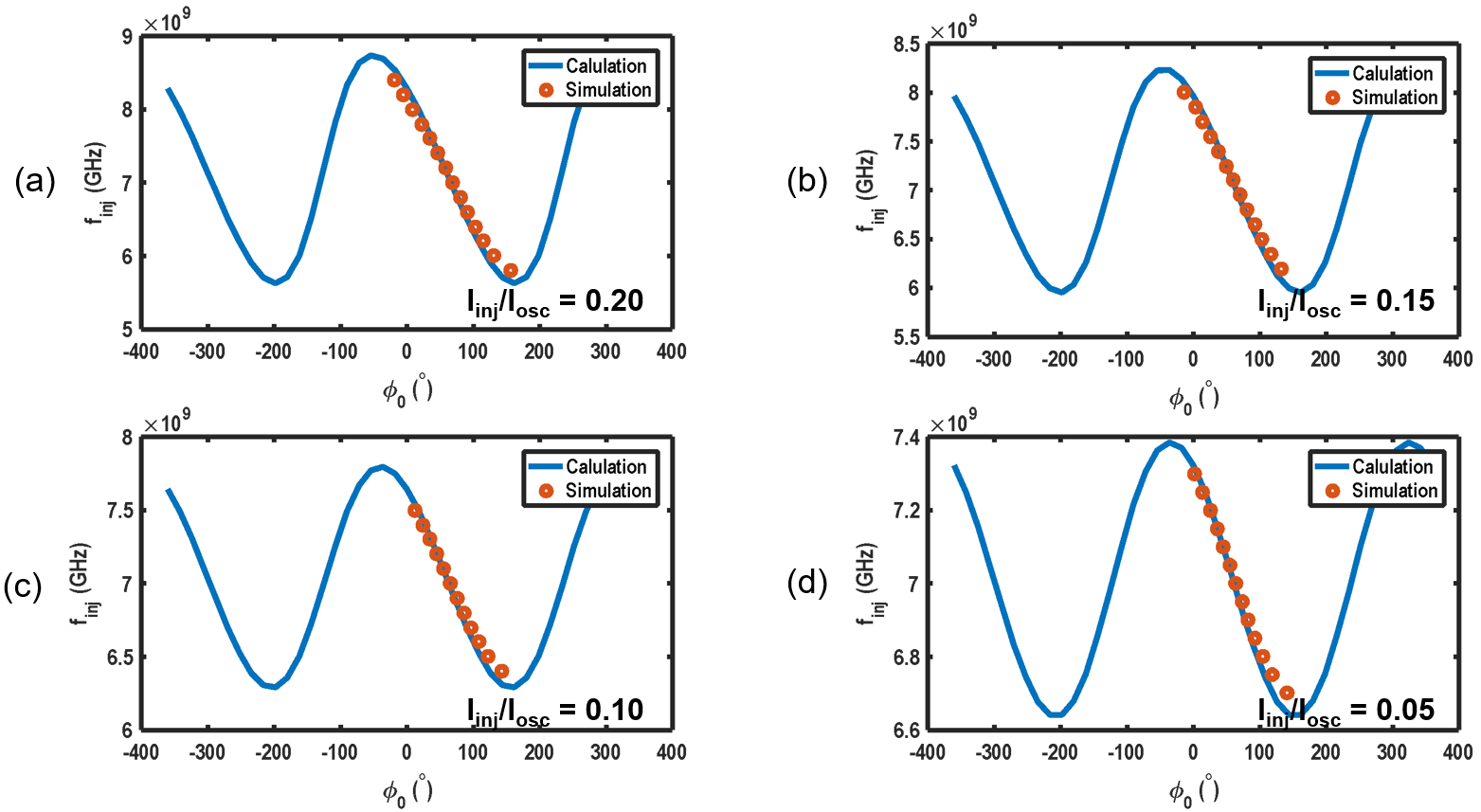}
\vspace{-6pt}
\caption{\small{Calculated and simulated $\phi_0$ versus injection frequency under injection strength: (a) 0.2; (b) 0.15; (c) 0.10; (d) 0.05.}}
\label{fig:sw_finj_phi0}
\vspace{-12pt}
\end{figure}

\begin{figure}[t!]
\centering 
\includegraphics[width=0.48\textwidth]{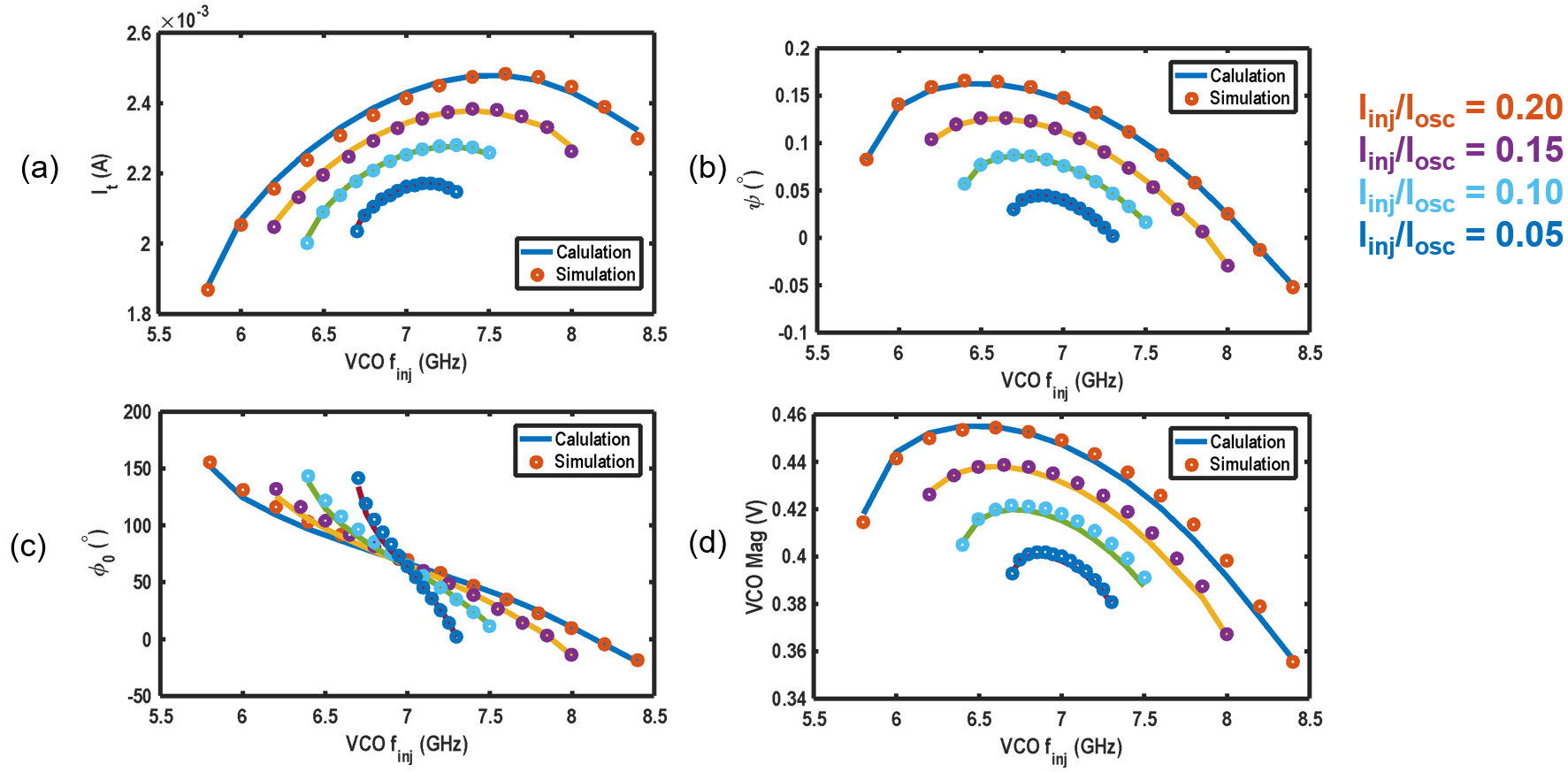}
\vspace{-6pt}
\caption{\small{Calculated and simulated $I_t$,$\psi$,$\phi_0$,$V_{osc}$ versus injection frequency under injection strength: (a) 0.2; (b) 0.15; (c) 0.10; (d) 0.05.}}
\label{fig:sw_finj_phi0_var}
\vspace{-12pt}
\end{figure}

Next, we fix the free-running frequency at 7 GHz and sweep the injection frequency under varying injection strengths of 0.05, 0.10, 0.15, and 0.20. As shown in Fig.~\ref{fig:sw_finj_phi0}, the equations provide analytical solutions for $\phi_0$, corresponding to specific locking ranges for each injection strength. The results demonstrate a wider locking range at the high-frequency end and a narrower range at the low-frequency end. Fig.~\ref{fig:sw_finj_phi0_var} compares the current amplitude, $\phi_0$ and $\psi$, and oscillation amplitude across the injection frequency for different injection strengths. The extended equations accurately predict these values, further validating the amplitude-to-phase conversion effect in injection-locked oscillators.


\section{Conclusions}
\label{sec:conclusions}
This paper presents an enhanced analytical model that extends Adler’s equation by accounting for amplitude-to-phase conversion. Simulation results show excellent agreement with the analytical predictions, confirming the model’s accuracy in capturing the locking range and phasor dynamics. The proposed equations effectively describe the phase and amplitude relationships in injection-locked ring oscillators and can be further extended to predict phase noise sensitivity through small-signal analysis.



\bibliographystyle{IEEEtran}
\bibliography{IEEEabrv,Journal_ILO}
\end{document}